\RequirePackage{fix-cm}

\documentclass[smallextended]{svjour3}       
\smartqed  
\usepackage{graphicx}
\usepackage{amsmath}
\usepackage{amssymb}
\usepackage{multirow}
\usepackage{epstopdf}
\usepackage{lscape}
\usepackage{adjustbox}
\usepackage{verbatim}

\begin{document}

\title{Low Complexity Reconfigurable Modified FRM Architecture with Full
Spectral Utilization for Efficient Channelizers}

\author{ Parvathi A. K.*\and  V. Sakthivel}

\titlerunning{}
\institute{Parvathi A. K\\
			  Department of Electronics and Communication Engineering\at
              National Institute of Technology Calicut\\
              Kerala, India\\
              Tel.: +91-9074555938\\
              \email{parvathi$_{-}$p180072ec@nitc.ac.in}           
           \and
           V. Sakthivel \at
              Department of Electronics and Communication Engineering\at
              National Institute of Technology Calicut\\
              Kerala, India\\
              Tel.: +91-9995335962\\
              \email{sakthi517@nitc.ac.in} 
}

\date{Received: date / Accepted: date}

\maketitle
\begin{abstract}
This paper proposes a design of low complexity, reconfigurable and narrow transition band (NTB) filter bank (FB). In our proposed Modified Frequency Response Masking (ModFRM) architecture, the modal filter and complementary filter in conventional FRM approach are replaced by a power complementary and linear phase FB. Additionally, a new masking strategy is proposed by which an M-channel FB can be designed by alternately masking even channels and odd channels. By using this masking strategy, it is only necessary to alternately mask even channels and odd channels by the two masking filters while modulating them over
the multiple spectra replicas appearing in [0,2$\pi$]
to generate uniform ModFRM FB. Also, reconfigurability of filter bank can be achieved
by adjusting the interpolation values appropriately, to obtain
more masking responses. To reduce the overall implementation complexity, masking filters are optimized using the interpolated FIR (IFIR) technique. The results indicate that the proposed method requires substantially less multipliers  in comparison to the reconfigurable FB existing in literature. Finally, non-uniform FB
are generated from the uniform FB by combining nearby channels. The proposed non-uniform ModFRM FB is used for the extraction of different
communication standards in the software defined radio (SDR)
channelizer. When more channels are to be extracted, the proposed scheme was able to achieve a reduced hardware complexity in comparison to other filter bank based SDR channelizers.

\end{abstract}

\section{Introduction}
\label{intro}

\par With the outbreak of modern wireless communication technologies, the compulsion to use the scarcely available spectrum fully and efficiently is at its peak. Channelizers in SDRs
are one of those applications that requests for a sharp digital filter of extremely narrow transition band (NTB) with higher stop band attenuation  for an alias-free switching among the preferred frequency bands with minimal adjacent channel interference.

\par The Frequency Response Masking (FRM) approach offers an enticing alternative among the
research efforts undertaken over the past few years to develop
such sharp filters with reduced hardware complexity. The underlying
idea of the FRM technique is from the interpolated FIR
(IFIR) technique proposed in \cite{neuvo1984interpolated}, which decomposes the sharp narrow pass band filter into a periodic model filter and an image
suppressor. However, as IFIR technique reduces the passband
width along with transition width, it is appropriate only
for narrow passband design. The beauty of FRM technique is
that it breaks the higher order filter designs into a group of filters
with reduced design requirements. The FRM method promises lesser complexity in comparison to the minimax optimum filter of infinite word length while meeting similar requirements. This is owing to the relatively lesser percentage of nonzero coefficients in the FRM scheme.

\par Rodrigues came up with a completely different FRM approach for the design of a linear phase sharp FIR filter by using a bandpass filter in \cite{rodrigues2005modified}. This method removed one of the masking filters and  interpolated filter from the conventional FRM technique and thus provides less complexity. For the development of arbitrary bandwidth sharp FIR filters , FRM and single filter frequency masking (SFFM) techniques are combined in \cite{yang2006new}. The final filter is formulated by a single masking filter with significantly lower arithmetic operations at the price of additional delays. A revised FRM strategy in which the IFIR technique implements one of the subfilters is discussed in \cite{lian2001improved}. 
 In \cite{wei2010frequency}, a serial masking scheme is implemented to perform the two-stage masking process, attempting to reduce masking filter complexity. The arithmetic complexity is reduced in \cite{saramaki2002optimization}
by using a common filter part to construct the masking filters in the traditional FRM structure. This technique requires swapping between the common filter portion and one of the filters for masking. This approach needs a relatively large number of switches to realize more channels and thus it becomes an arduous task to reduce the complexity of switching.

\par Nevertheless, due to the limitations of the FRM architecture, the design of the FRM FB includes complicated modulations of interpolated filters and masking filters. Also, each time the specification of the desired filter response is changed, masking filters must be redesigned. This is troublesome in applications involving a bank of filters such as multi-carrier systems, generating matched filters for radar target signatures, etc. Up to now, an efficient realization of flexible and NTB FB is still a challenging work in multi-standard wireless communication applications. 

\par This paper proposes a design of low complexity, reconfigurable NTB FB. The key contributions of this work can be outlined as follows.
(1) In our proposed Modified FRM (ModFRM) architecture, the modal filter and complementary filter in conventional FRM approach are replaced by a power complementary and linear phase FB. (2) An alternate masking scheme is introduced in the second phase for realizing the uniform ModFRM FB which guarantees full spectral utilization with reduced implementation complexity. (3) Compared to previous approaches, it is only necessary to alternately mask even channels and odd channels by the two masking filters while modulating them over
the multiple spectra replicas appearing in [0,2$\pi$]
to generate uniform ModFRM FB. This concept of ModFRM FB that takes advantage of
all the images resulting from the interpolation process, just by alternately masking with the two  masking filters is indeed a novel one.  To lower the multiplicity of the masking filters even further, IFIR technique is adopted. (4) Reconfigurability of the proposed method is claimed by designing different ModFRM FB from
the same modal filters, by changing the interpolation factor, with a very small increase in multipliers required. Results prove that compared to other reconfigurable FB existing in literature, the proposed ModFRM FB ends in a substantial drop in the number of multipliers. (5) Finally, a reconfigurable SDR channelizer is designed by combining nearby channels of uniform ModFRM FB. 

\par The paper is structured as follows: Section 2 gives a brief description about existing FRM
technique. In section 3, together with the basic ideas that laid the foundation for the structure, the proposed ModFRM based NTB filter is discussed. Section 4 illustrates an alternate masking scheme proposed for generating the ModFRM FB. Section 5 includes the design examples and result analysis. Section 6 concludes the paper.

\section{Review of Frequency response masking technique}

\par
\begin{figure}
	\centering
	\includegraphics[width=0.6\linewidth,height=2cm]{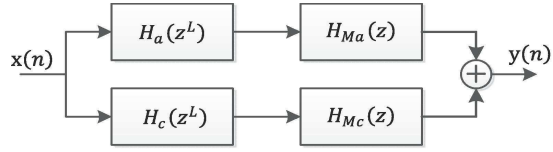}
	\caption{FRM structure proposed by Lim.\cite{lim1986frequency}}
	\label{frm}
\end{figure}

 The beauty of the FRM technique \cite{lim1986frequency} is that it breaks the higher-order ﬁlter design into the design of four sub-filters, namely, the modal ﬁlter denoted by $H_{a}(z)$, complementary ﬁlter denoted by $H_{c}(z)$, masking ﬁlter $H_{Ma}(z)$ and complementary masking ﬁlter $H_{Mc}(z)$ which has reduced design requirements. Fig. \ref{frm} demonstrates the basic structure of the FRM \cite{lim1986frequency}. The interpolated model filter also known as band-edge shaping filter and its complementary filter together constructs the NTB filter of arbitrary passband. By cascading the two masking filters $H_{Ma}(z)$ and $H_{Mc}(z)$ with $H_{a}(z)$  and $H_{c}(z)$, the unnecessary passbands are eliminated.

\par
\begin{figure}
\centering
	\includegraphics[width=0.65\linewidth,height=7.7cm]{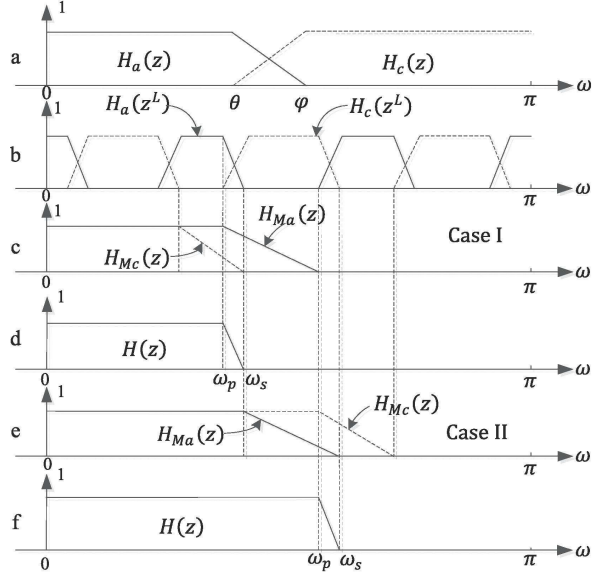}
	\caption{Illustration of the FRM structure proposed by Lim.\cite{lim1986frequency}}
	\label{illustrate_frm}
\end{figure}

 Illustration of the FRM structure is demonstrated in Fig. \ref{illustrate_frm}. There are two distinct cases: Case I and Case II. $H_{a}(z^{L})$  and $H_{c}(z^{L})$  determine the frequency response of the overall filter $H(z)$, near the transition band, in case I and case II respectively.
The transfer function of the FRM filter is given by,
\begin{equation}
H(z)=H_{a}(z^{L})H_{Ma}(z)+ H_{c}(z^{L})H_{Mc}(z)
\label{FRM}
\end{equation}
where $H_{c}(z)$ can be realized as shown:
\begin{equation}
H_{c}(z)=z^{-(N-1)/2}-H_{a}(z)
\label{complementaryfilter}
\end{equation}
Here N is the length of the filter $H_{a}(z)$ and \textit{L} is the interpolation factor. 
\section{Proposed Modified FRM (ModFRM) structure}

\par The fundamental ideas which paved the way for the proposed ModFRM structure can be elaborated as follows. 
\par It is well known that modulation can convert a low-pass filter to a set of band pass filters. If the passband and transition width of the low pass model filter is chosen appropriately, then its \textit{M} modulated filters can completely cover the entire spectrum which is the idea utilized in this work.

\par The number of passbands $N_{p}$, formed by interpolating $H(z)$ by \textit{L} can be expressed as:
\begin{equation}
N_{p}=\left \lceil M+1 \right \rceil
\label{Np}
\end{equation}
\par On interpolating by \textit{L}, the bandwidth of the \textit{i}th passband denoted by $B(i)$, can be defined as follows:

\begin{equation}\label{Bi}
B(i)=\left\{\begin{matrix}
& \frac{B_{0}}{L};   \;\;\;\;\;\;     i=1 \;and \;N_{p}\\ 
& \frac{2B_{0}}{L};  \;\;\;\;\;\;      otherwise
\end{matrix}\right.
\end{equation}

where $B_0$ is the bandwidth of the initial filter \textit{H(z)}. The center frequencies of the passbands, $\omega _{k}$ , can be determined by,
\begin{equation}\label{omegak}
\omega _{k}=\frac{2\pi}{L}k;  \;\;\;\;\;\;  k=0,1.....N_{p}-1
\end{equation}
\begin{figure}
	\centering
	\includegraphics[width=0.85\linewidth,height=4.85cm]{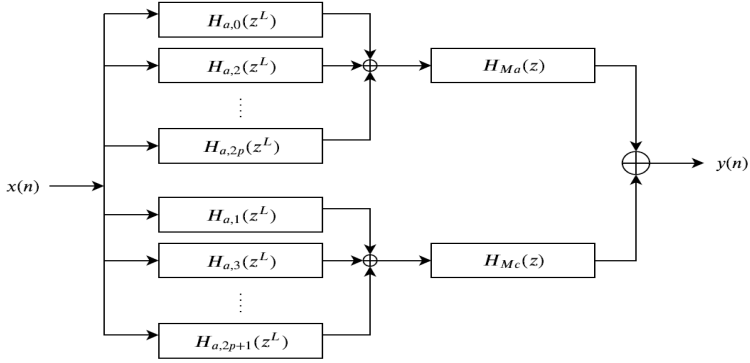}
	\caption{Proposed ModFRM structure.}
	\label{ModFRM}
\end{figure}

\par By changing \textit{L}, it is possible to change the number and locations of $H(z^{L})$ passbands. In addition, from Eq. (\ref{Bi}), we can see that the bandwidths of the $H(z^{L})$ passbands can also be modified through varying the \textit{L}. 
The above features of the interpolated filters laid the foundation of the proposed filter bank reconfigurability.
\par The layout of the ModFRM is as in Fig. \ref{ModFRM}. The modal and complementary filters in traditional FRM structure are replaced by interpolated odd and even modulated DFT channels of the modal filter. The unwanted bands in the interpolated modulated filter bank are then eliminated by the cascade connection of the two masking filters to produce the desired NTB filter, in harmony with the basic principle of FRM. 
\par
\begin{figure}
	\centering
	\includegraphics[width=\linewidth, height=9cm]{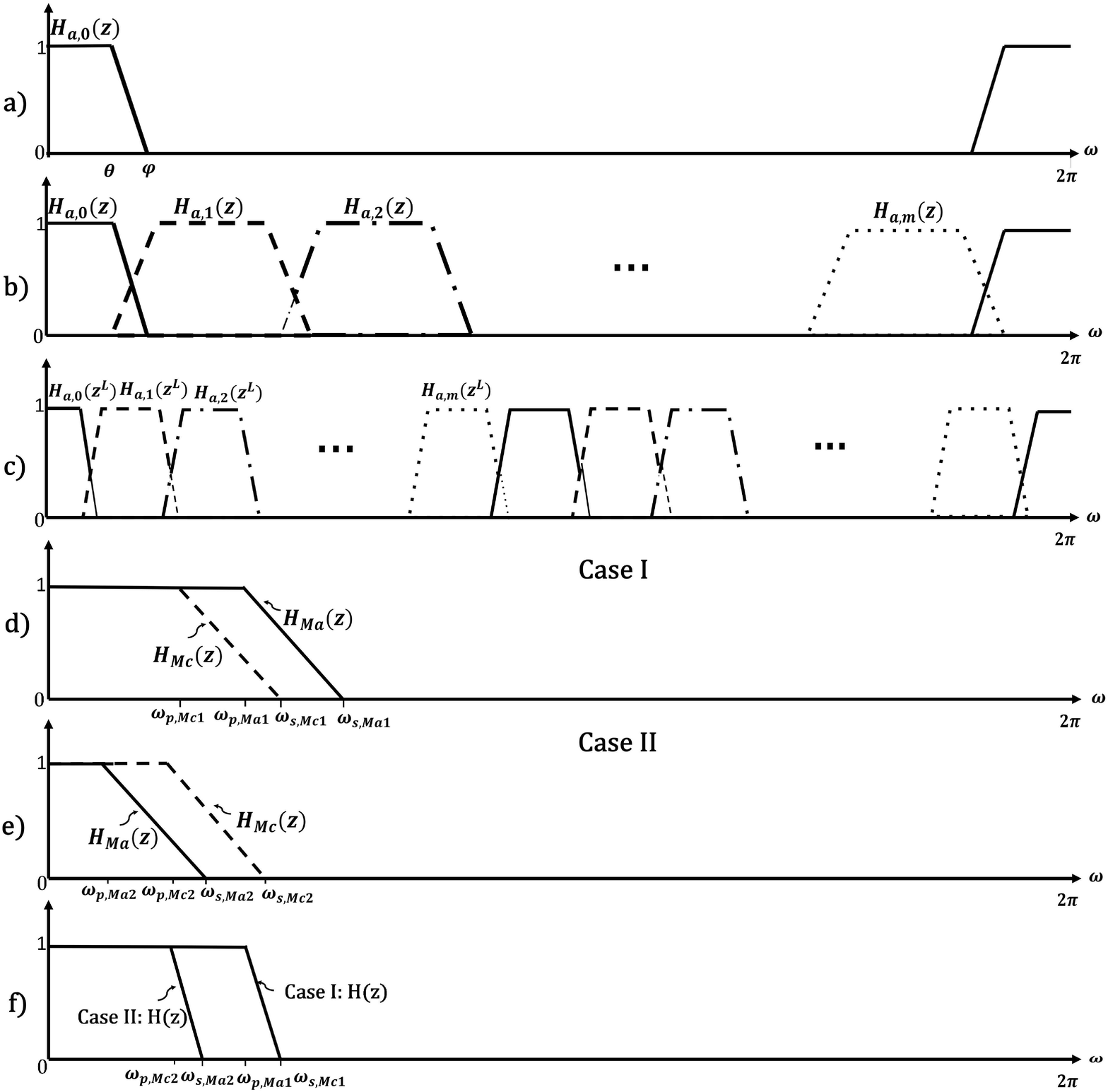}
	\caption{Illustration of the proposed ModFRM structure.}
	\label{llustrate_MoFRM}
\end{figure}

 As shown in Fig. \ref{llustrate_MoFRM}a and Fig. \ref{llustrate_MoFRM}b, the low pass linear phase model FIR filter  $h_{a,0}(n)$ of length $N_{a}$ is complex modulated to obtain all the analysis filters. This low pass filter needs to have a suitable passband and transition band to ensure that the modulated filters can cover the entire spectrum.
Let $\theta$ and $\phi$ be the passband and stopband edge frequencies of the modal filter.\\

\textbf{\textit{The necessary conditions for the design of ModFRM FB:}}\
\begin{enumerate}
 \item For the modal filter, it is necessary to select the appropriate passband and transition band to ensure that the modulated filters can cover the entire spectrum.
 \begin{equation}\label{2pi}
	(m+1)(\theta+\phi)=2\pi
	\end{equation}
	\item The number of modulated filters m must be an odd number to guarantee that when we recombine the modulated filters, the frequency response appears alternately.
	\begin{equation}
	m=2p+1, p=0,1,2...
	\end{equation}
	 \item Additionally, the transition band of the modal filter needs to satisfy the power complementary condition. This guarantees that the interpolated odd and even modulation channels will form a power complementary and linear phase FB.
\end{enumerate}

\par The center frequency of the q-th modulated channel filter is given by,
\begin{equation}
\omega _{q}=\pi q(\theta +\phi );\;\;where\;\; q\;= 1\;to\;m
\end{equation}  
and $h_{a,q}(n)$ can be expressed as follows:
\begin{equation}
h_{a,q}(n)=h_{a,0}(n)e^{j\omega _{q}n }
\end{equation}
\par Let $H_{a,q}(e^{j\omega})$ represents the frequency response of the q-th channel modulated filter $h_{a,q}(n)$. Then $H_{a,q}(z)$ is given by,
\begin{equation}
H_{a,q}(z)=\sum_{n=0}^{N_{a}-1}h_{a,q}(n)z^{-n}=\sum_{n=0}^{N_{a}-1}h_{a,0}(n)e^{j\omega _{q}n }z^{-n}
\end{equation}
\par Other modulated FBs in literature, like modified DFT,
having a similar cutoff frequency condition as that of DFT modulated FB for the initial low pass filter can
also be considered. Each of the modulated filters are then interpolated by a factor \textit{L}, which reduces its transition width by the same factor. $ H_{a,q}(z^{L}) $ represents the interpolated q-th DFT modulated filter which is depicted in Fig. \ref{llustrate_MoFRM}c.  The two sets of masking filters are shown in Fig. \ref{llustrate_MoFRM}d and Fig. \ref{llustrate_MoFRM}e. \par There are two different design cases: Case I and Case II in which the frequency response of the overall filter, $H(z)$ near the transition band is determined by the sum of interpolated even modulated channels (called as even band edge shaping filter) and the sum of odd modulated channels (called as odd band edge shaping filter) respectively. According to \cite{lim1986frequency}, passband and stopband edge frequencies of masking filters  can be derived as follows for the cases I and II:
\begin{align}
\begin{tabular}{lr}
Case I \\
\hline \\
$\begin{aligned}[t] 
\omega _{p,Ma1}=\frac{3\theta+2\phi}{L}; \omega _{s,Ma1}=\frac{4\theta+3\phi}{L};\\ \\
\omega _{p,Mc1}=\frac{\theta+2\phi}{L}; \omega _{s,Mc1}=\frac{2\theta+3\phi}{L};\\
\end{aligned}$ \\ \\
\hline
\hline
Case II: \\
\hline\\
$\begin{aligned}[t] 
\omega _{p,Ma2}=\frac{\phi}{L}; \omega _{s,Ma2}=\frac{\theta+2\phi}{L};\\ \\
\omega _{p,Mc2}=\frac{2\theta+\phi}{L};\omega _{s,Mc2}=\frac{3\theta+2\phi}{L};\\
\end{aligned}$ \\ \\
\hline
\end{tabular}
\end{align}

\par The complexity of both masking filters depends on the \textit{L} and the cutoff frequencies of the band-edge shaping filter. The sum of the two masking filters transition widths is equal to $\frac{1}{L}$ as in \cite{lim1986frequency}. Because of this constraint, it is difficult for the FRM technique to increase L and also keep the complexity of the masking filters within a certain limit. If L is very large it results in denser replicas in the frequency response of the shaping filters, which in turn results in sharp masking filters. Hence masking filters are optimized using the IFIR technique with an interpolation factor, $L_{IFIR}$ which gives the lowest complexity. 
The optimum $L_{IFIR}$, that minimizes the total number of multiplications, can be evaluated by considering the lengths of all sub-filters for various $L_{IFIR}$ and deciding the best-case outcome heuristically. Masking filters can also be implemented very efficiently using a polyphase structure, as the coefficient values of band-edge shaping filters are non zero at intervals of \textit{L} samples.
\par The overall transfer function of the proposed ModFRM filter can be expressed as:
\begin{align}\label{eq:overallTF}  
H(z)=[ H_{a,0}(z^{L})+H_{a,2}(z^{L})+...+H_{a,2p}(z^{L}) ]H_{Ma}(z)\nonumber\\
+[H_{a,1}(z^{L})+H_{a,3}(z^{L})+...+H_{a,2p+1}(z^{L})]H_{Mc}(z)
\end{align}
The desired NTB filters obtained in case I and II are shown in Fig. \ref{llustrate_MoFRM}f. 
\par In order to make use of remaining multiple spectra replicas appearing in [0, $2\pi$], separate masking filters have to be designed which is straightforward and tremendous. So as to reduce the complexity, the masking filters  had better be reusable. In light of condition (\ref{omegak}), as long as the \textit{L} is not varied, the  passband width and locations of $H_{a,q}(z^{L})$ are fixed. This suggests that passbands generated by various interpolated filters that cover the entire frequency range can share the same set of masking filters. This idea inspired for the alternate masking scheme proposed in the next section.
\section{Alternate Masking Scheme for ModFRM FB}
\begin{figure}
 	\centering
	\includegraphics[width=\linewidth, height=9cm]{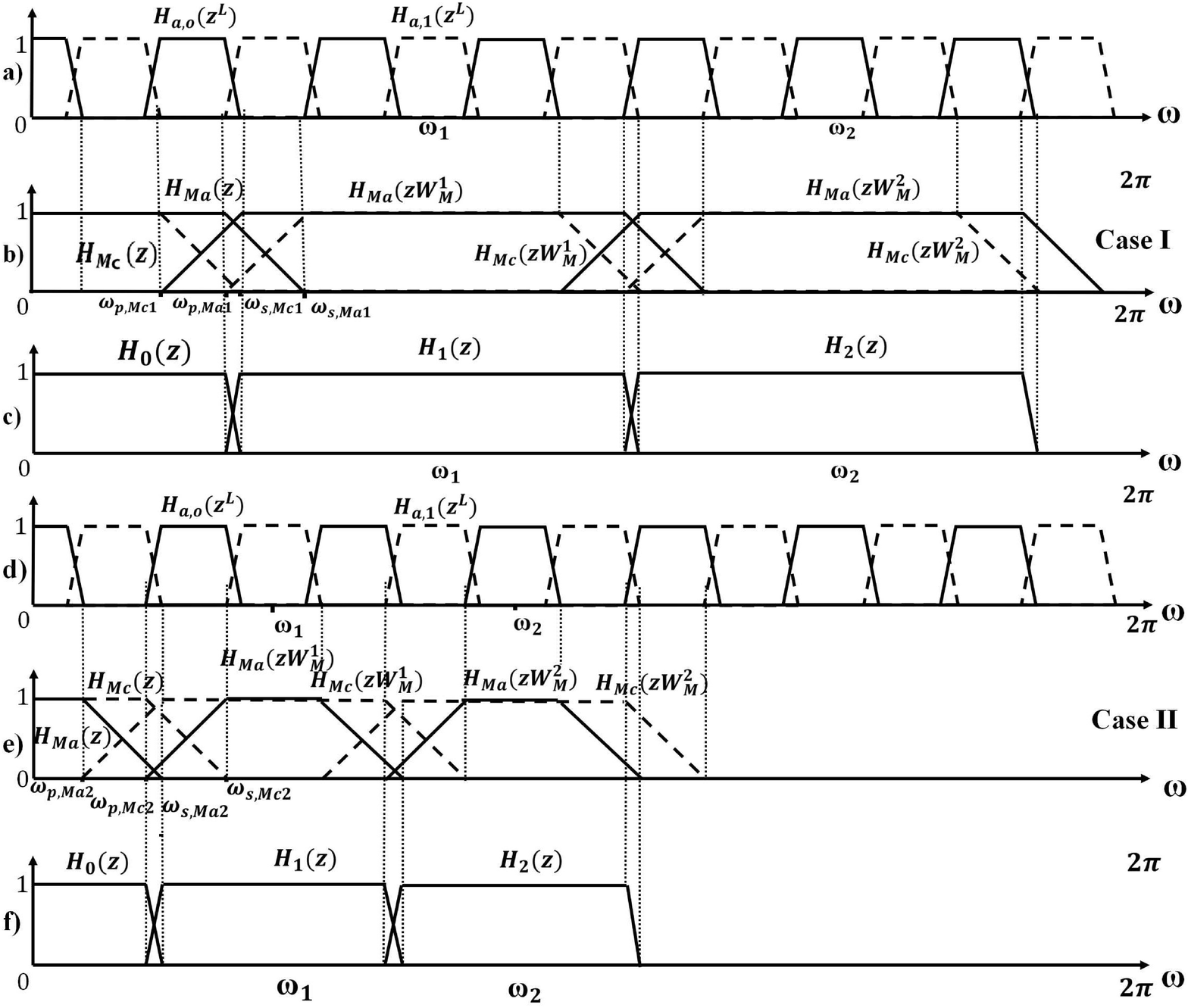}
	\caption{Illustration of the proposed alternate masking scheme.}
	\label{fig:AMS}
\end{figure}

The proposed alternate  masking strategy for ModFRM FB generation is shown in Fig. \ref{fig:AMS}. Here the masking filters are DFT modulated and passed over the remaining multiple spectra replicas appearing in [0, $2\pi$] to generate uniform ModFRM FB. Unlike the conventional FRM, modulated masking filters $H_{Ma}(z{W_{M}^{k}})$ and $H_{Mc}(z{W_{M}^{k}})$ alternately masks the odd and even channels, respectively. To simplify the analysis, the number of modulated filters is taken as \textit{m}=1. As seen in Fig. \ref{fig:AMS}a, $H_{a,0}(z^{L})$ and $H_{a,1}(z^{L})$ represent  the interpolated modal filter and DFT modulated filter, respectively.
As shown in Fig. \ref{fig:AMS}b, $H_{Ma}(z)$ and $H_{Mc}(z)$ masking filters are used respectively to mask $H_{a,0}(z^{L})$ and $H_{a,1}(z^{L})$. Then the FRM filter $H_{0}(z)$ can be obtained as Fig. \ref{fig:AMS}c. But the situation is different in the second channel. $H_{a,0}(z^{L})$ should be masked with $H_{Mc}(z{W_{M}^{1}})$ instead of $H_{Ma}(z{W_{M}^{1}})$. Similar is for case II which is shown in Fig. \ref{fig:AMS}e and \ref{fig:AMS}f.
\par Finally, the analysis filter obtained can be represented as in (\ref{eq:overallTF}).
\begin{align}\label{eq:overallTF}  
H_{k}(z)=\sum_{i=0}^{p}H_{a,2i}(z^{L})H_{Ma}(z{W_{M}^{k}})\nonumber\\ +\sum_{i=0}^{p}H_{a,2i+1}(z^{L})H_{Mc}(z{W_{M}^{k}})\;\;\;\;\;;for\;\;k\;\;even\nonumber\\
H_{k}(z)=\sum_{i=0}^{p}H_{a,2i}(z^{L})H_{Mc}(z{W_{M}^{k}})\nonumber\\+ \sum_{i=0}^{p}H_{a,2i+1}(z^{L})H_{Ma}(z{W_{M}^{k}})\;\;\;\;\;;for\;\;k\;\;odd
\end{align}
where $W_{M}=e^{-j\frac{2\pi}{M}}$,k=0,1,...M-1.\\
\par The uniqueness of our proposed architecture is that by using common masking filters it can efficiently exploit the redundancy in multi-stage realization. Reconfigurability can be accomplished by adjusting L appropriately, result in variable bandwidth uniform FB.
\par For the desired filter with the stopband frequency $f_s$, maximum number of channels is found as:
\begin{equation}\label{max_channels}
C_{max}=\left \lceil \frac{1}{f_{s}} \right\rceil   \end{equation}
The number of channels, \textit{C} that can be obtained is dependent on the number of modulations \textit{m} and the interpolation factor \textit{L} as follows:\\

\begin{equation}
C=\left\{\begin{matrix}
& \frac{(m+1)\times L}{5};   \;\;\;\;\;\;    for\;case\; I\\ \\
& \frac{(m+1)\times L}{3};  \;\;\;\;\;\;     for\; case\; II
\end{matrix}\right.
\label{casei_ii}
\end{equation}

\par The various wireless protocols, however, have different
channel spacing or bandwidth, and therefore necessitate non-uniform
subband decomposition. The non-uniform FB decomposes the input signal into uneven bandwidth subbands that are derived by combining the adjacent channels from the uniform FB \cite{sakthivel2018low}. The analysis filter  $H_{i}(z)$  produced by combining $a_{i}$ neighboring analysis filters can be represented as below:
\begin{align}
\tilde{H_{i}(z)}=\sum_{k=n_{i}}^{n_{i}+a_{i}-1}H_{k}(z),\;\;\;\;\;i=0,1...M-1
\label{nonuni}
\end{align}
where $n_{i}$ is the channel number.
\par By multiplying with a sign-change block S, one can obtain linear phase non-uniform FB using the method suggested in \cite{lee1995design}. For example, for the implementation of the 5-channel NU ModFRM FB from an 8 channel uniform ModFRM FB with the channel allocation ($c_{0}$,$c_{1}$,$c_{2}$,$c_{3}$)=(2,1,3,2), sign change block can be expressed as:
\begin{equation*}
S =
\begin{bmatrix}
1 & 1 & 0 & 0 & 0 & 0 & 0 & 0 \\
0 & 0 & 1 & 0 & 0 & 0 & 0 & 0\\
0 & 0 & 0 & 1 & 1 & 1 & 0 & 0\\
0 & 0 & 0 & 0 & 0 & 0 & 1 & 1
\end{bmatrix}
\end{equation*}
\par There are no severe conditions as in \cite{sakthivel2018low} for combining adjacent channels. Hence any adjacent channels can be combined to generate NU ModFRM FB.

\section{Results and discussion}
\par Two examples are presented in this section, to show the low complexity and reconfigurability properties
of the ModFRM FB, and its application in SDR channelizers. Example 1 illustrates the generation of  different ModFRM FB from
the same modal filter by varying the \textit{L}. Example 2 demonstrates the design of
reconfigurable ModFRM FB for SDR channelizers. Parks McClellan algorithm is used to design the filters in our work. The interpolation is attained by placing L delay elements instead of each delay element. As multiplication in a digital FB is the most complicated and power-consuming task, this section offers a comparison of the real-valued multipliers required in the proposed FB with other FB's existing in the literature. One should also note that the blocks of modulation also add to the FB overall complexity. Nonetheless, this contribution is independent of the filter orders and has a fairly low complexity relative to the filter portion \cite{vaidyanathan2006multirate,rosenbaum2006approach}. Therefore, it is not being addressed here.\\
\subsection{Example 1}
In this example, different ModFRM FB are designed from the same modal filter, by varying the \textit{L}. The specifications of the modal filter are chosen as  passband and stopband frequencies $0.2\pi$ and $0.3\pi$, stopband attenuation as 60dB, and passband ripple as 0.0065dB. The passband and stopband edge frequencies are iteratively adjusted to satisfy the 3-dB requirement, retaining the transition width constant.

\begin{figure}
	\centering
	\includegraphics[width=0.85\linewidth,height=4.25cm]{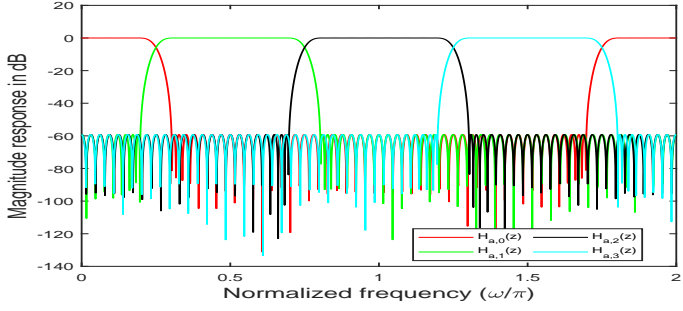}
	\caption{Frequency response of modal filter $H_{a,0}(z)$ and its modulated sub-filters. }
	\label{fig:proto_filter}
\end{figure}
\par In the design approach of the FB, the transition width of the
8-channel FB prototype filter differs from that of the
32-channel, as minimal amplitude distortion and aliasing
has to be assured. As the spectrum occupied by the 8 filter bandwidths in 8-channel FB has to be occupied by the 32 filter
bandwidths in 32-channel FB, the transition width required for the 32-channel FB prototype filter is sharper than the 8-channel FB for minimal aliasing.
In the case of 8-channel FB, the normalized transition width of prototype filter is taken as 0.01. The normalized transition
width must be sharper for a 32-channel FB and it is taken as 0.01$/$4 = 0.0025.
\begin{figure}
	\centering
	\includegraphics[width=0.85\linewidth,height=4.25cm]{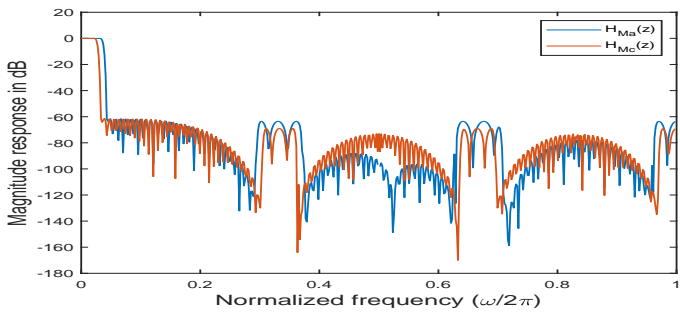}
	\caption{Masking filters $H_{Ma}(z)$ and $H_{Mc}(z)$ designed using IFIR technique.}
	\label{fig:mk_filter}
\end{figure}
\begin{table*}[]
	\centering
	\caption{FRM filter parameters for different values of L $\&$  corresponding total number of multipliers.}
	\begin{adjustbox}{width=\linewidth}
		\small
	\begin{tabular}{ccccccccccccc}
\hline
\multirow{2}{*}{\begin{tabular}[c]{@{}c@{}}Interpolation\\  factor, L\end{tabular}} & \multirow{2}{*}{\begin{tabular}[c]{@{}c@{}}No. of \\ channels\end{tabular}} & \multicolumn{3}{c}{$H_{Ma}(z)$}                                                       & \multicolumn{3}{c}{$H_{Mc}(z)$}                                                         & \multirow{2}{*}{$L_{IFIR}$} & \multicolumn{3}{c}{$H(z)$}                                                             & \multicolumn{1}{l}{\multirow{2}{*}{\begin{tabular}[c]{@{}l@{}}Total no. of multipliers\\ $(M_{modal},M_{ma,pr},M_{ma,is},M_{mc,pr},M_{mc,is})$\end{tabular}}} \\ \cline{3-8} \cline{10-12}
                                                                                    &                                                                             & Pb         & Sb          & \begin{tabular}[c]{@{}c@{}}Transition\\ width\end{tabular} & Pb          & Sb          & \begin{tabular}[c]{@{}c@{}}Transition \\ width\end{tabular} &                             & Pb         & Sb          & \begin{tabular}[c]{@{}c@{}}Transition \\ width\end{tabular} & \multicolumn{1}{l}{}                                                                                                                                          \\ \hline
10                                                                                  & 8                                                                           & 0.12$\pi$  & 0.17$\pi$   & 0.05$\pi$                                                  & 0.08$\pi$   & 0.13$\pi$   & 0.05$\pi$                                                   & 3                           & 0.12$\pi$  & 0.13$\pi$   & 0.01$\pi$                                                   & (32+25+9+24+8)=98                                                                                                                                             \\
15                                                                                  & 12                                                                          & 0.08$\pi$  & 0.1132$\pi$ & 0.0333$\pi$                                                & 0.0535$\pi$ & 0.0868$\pi$ & 0.0333$\pi$                                                 & 4                           & 0.08$\pi$  & 0.0866$\pi$ & 0.0066$\pi$                                                 & (32+28+11+22+14)=107                                                                                                                                          \\
20                                                                                  & 16                                                                          & 0.06$\pi$  & 0.085$\pi$  & 0.025$\pi$                                                 & 0.04$\pi$   & 0.065$\pi$  & 0.025$\pi$                                                  & 6                           & 0.06$\pi$  & 0.065$\pi$  & 0.005$\pi$                                                  & (32+25+17+24+15)=113                                                                                                                                          \\
25                                                                                  & 20                                                                          & 0.048$\pi$ & 0.068$\pi$  & 0.02$\pi$                                                  & 0.032$\pi$  & 0.052$\pi$  & 0.02$\pi$                                                   & 7                           & 0.048$\pi$ & 0.052$\pi$  & 0.004$\pi$                                                  & (32+26+20+26+17)=121                                                                                                                                          \\
30                                                                                  & 24                                                                          & 0.04$\pi$  & 0.0567$\pi$ & 0.048$\pi$                                                 & 0.0267$\pi$ & 0.0433$\pi$ & 0.0166$\pi$                                                 & 8                           & 0.04$\pi$  & 0.0433$\pi$ & 0.0033$\pi$                                                 & (32+28+22+27+18)=127                                                                                                                                          \\
40                                                                                  & 32                                                                          & 0.03$\pi$  & 0.0425$\pi$ & 0.0125$\pi$                                                & 0.02$\pi$   & 0.0325$\pi$ & 0.0125$\pi$                                                 & 9                           & 0.03$\pi$  & 0.0325$\pi$ & 0.0025$\pi$                                                 & (32+33+23+32+17)=137                                                                                                                                          \\ \hline
\end{tabular}
	\end{adjustbox}
	\label{table:table1} 
\end{table*}

\par
\begin{figure}
	\centering
\includegraphics[width=0.85\linewidth,height=4.25cm]{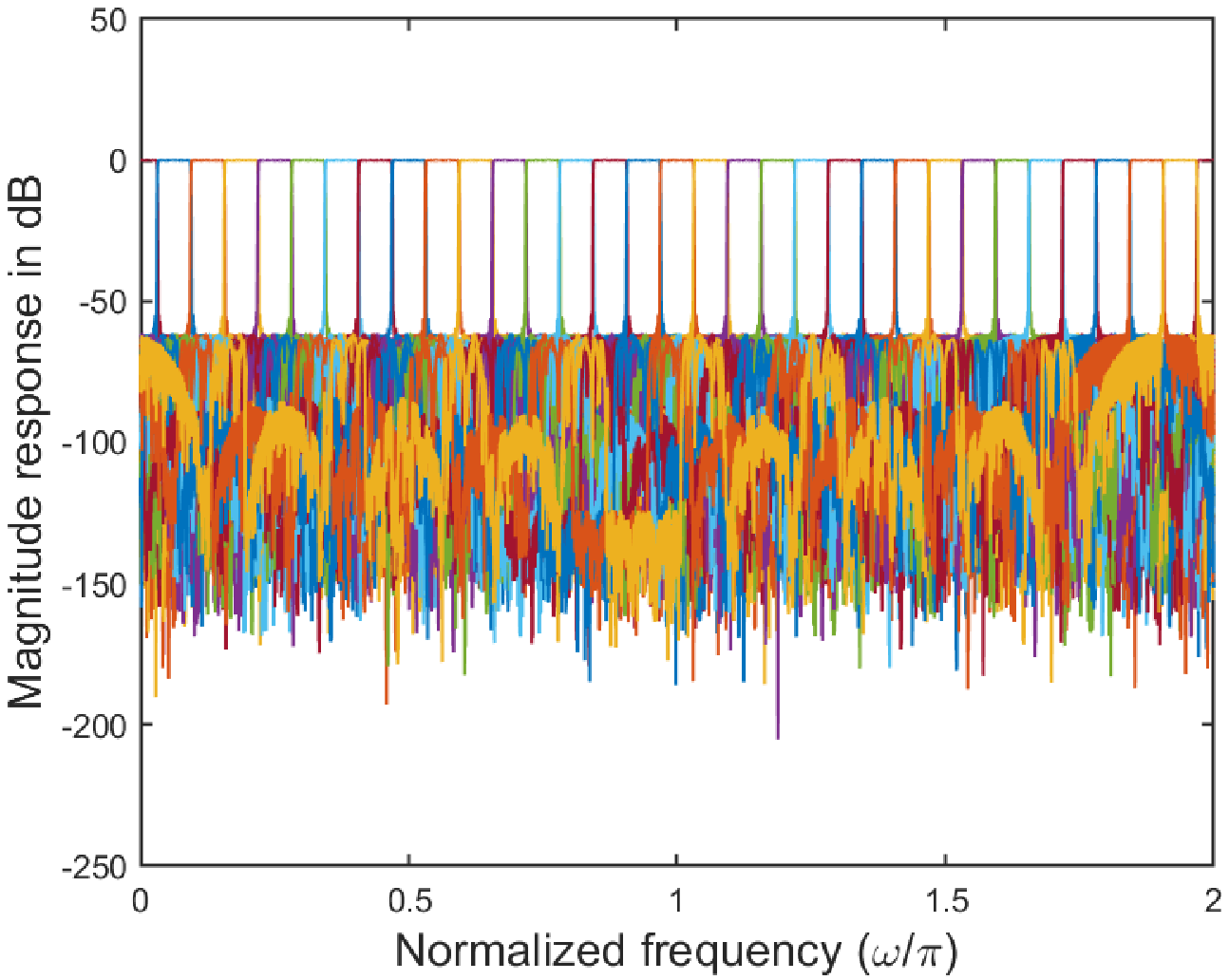}
\caption{Uniform 32 channel ModFRM FB.}
	\label{fig:M4L40C32}
	\end{figure}

\begin{figure}
	\centering
\includegraphics[width=0.85\linewidth,height=4.25cm]{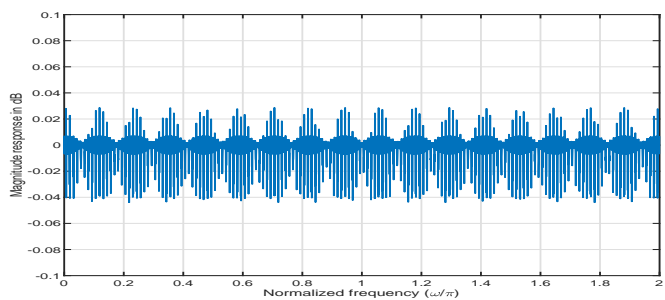}
	\caption{Amplitude distortion plot of uniform 32 channel ModFRM FB.}
	\label{fig:ampdist32}
	\end{figure}

 The DFT modulated filters for \textit{m}=3 and the masking filters designed using the IFIR technique when \textit{ L} is taken as 40 is shown in Fig. \ref{fig:proto_filter} and Fig. \ref{fig:mk_filter} respectively.  As per (\ref{casei_ii}), the 32 channels generated by the proposed alternate masking scheme is given in Fig. \ref{fig:M4L40C32}. The amplitude distortion plot of the filter bank is obtained as in Fig. \ref{fig:ampdist32}. Table \ref{table:table1} provides the sub-filter specifications for the different \textit{L} values. It can be noted that the modal filter is a fixed filter under all situations, and only the number of delays \textit{L} is modified. By masking with suitable masking filters and using the proposed alternate masking scheme, uniform ModFRM FB is obtained. The total number of multipliers is given by (\ref{eq:totalmultiplier}),
\begin{equation}
    M_{total}=M_{modal}+M_{ma,pr}+M_{ma,is}+M_{ma,pr}+M_{ma,is}
    \label{eq:totalmultiplier}
\end{equation}
where $M_{modal}$ is the number of multipliers in modal filter, $M{ma,pr}$ and $M_{ma,is}$ together constitutes the number of multipliers in the prototype filter and image suppressor filter which are the sub-filters of $H_{Ma}(z)$ designed using IFIR technique, $M{mc,pr}$ and $M_{mc,is}$ together constitutes the number of multipliers in the prototype filter and image suppressor filter which are the sub-filters of $H_{Mc}(z)$ designed using IFIR technique. Results in Table \ref{table:table1} show that the number of channels obtained can be varied by modifying the $\textit{L}$, with a very slight increase in multiplier numbers. A comparison of the number of multipliers for 32-channel FB achieved by different methods in literature is listed in Table \ref{table:table2}.

\begin{table}[]
	\centering
	\caption{Comparison of the number of multipliers}
	\begin{adjustbox}{width=\linewidth}
		\small
		\begin{tabular}{lccc}
			\hline
			\textbf{Method} & \textbf{No. of channels} & \textbf{Transition width} & \textbf{No. of multipliers} \\ \hline
			CMFB \cite{lee2006subband,zhang2018improved} & 32 & 0.0025$\pi$ & 2762 \\ 
			Improved FRM \cite{zhang2018improved} & 32 & 0.0026$\pi$ & 1140 \\ 
			MDFT \cite{sakthivel2018low} & 32 & 0.0025$\pi$ & 240 \\ 
			Proposed method & 32 & 0.0025$\pi$ & 137 \\ \hline\\
		\end{tabular}
	\end{adjustbox}
	\label{table:table2} 
\end{table}

\subsection{Example 2: SDR channelizer design using NU ModFRM FB}

\par The channelizer in an SDR must be adaptable to channels of different bandwidths in order to satisfy widely recognized wireless communication norms of different channel bandwidths as listed in Table 2 in \cite{sakthivel2018low}. Therefore, non-uniform filter bank would be better suited to use in the channelizer than uniform filter bank. Since the
channelizer performs at the highest sampling rate in the
receiver’s digital front end, low complexity, low power and
reconfigurability of the architecture are some of its essential
requisites. Ideally, the reconfigurability needs to be achieved
in such a way that the filter bank architecture supporting
the existing communications standard has to be remodelled
to accommodate a different communication standard with the
minimal overhead.

\begin{figure}
	\centering
	\includegraphics[width=0.85\linewidth,height=4.25cm]{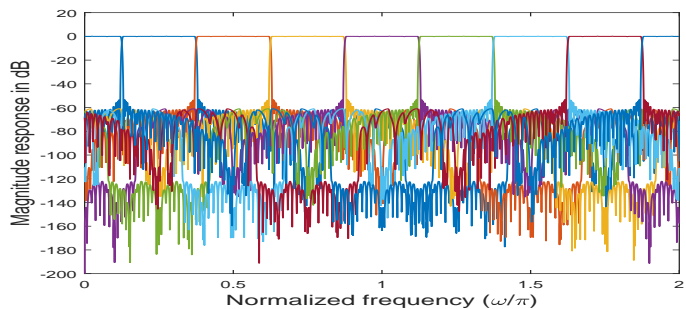}
	\caption{Uniform 8 channel ModFRM FB }
	\label{C8U}
\end{figure}

\begin{figure}
	\centering
	\includegraphics[width=0.85\linewidth,height=4.25cm]{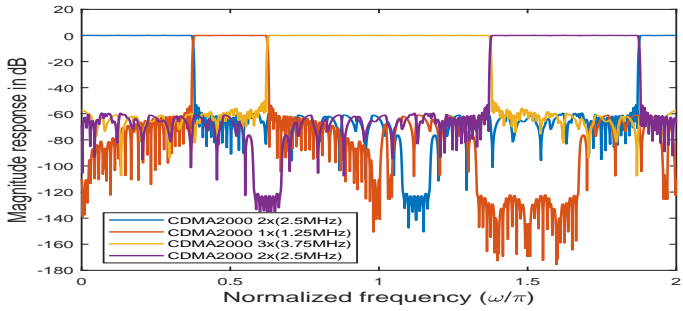}
	\caption{NU 4-channel ModFRM FB (for CDMA 2000) derived from uniform 8-channel ModFRM FB.}
	\label{C8NU4}
\end{figure}
\par The literature suggest distinct SDR channelizer designs using filter banks. Of all the alternative ideas by which channelization is realized, the use of Frequency response masking (FRM) method in SDR channelizer is an extensively investigated topic in recent research papers. 
A reconfigurable architecture of FRM filters is built in \cite{mahesh2008reconfigurable} to implement low power SDR channel filters. The FRM FB proposed in \cite{bindiya2012design} is more appropriate for single channel extraction than multiple channel extraction due to the necessity of separate masking filters to generate each channel. Channelizers are designed in \cite{mahesh2011filter} based on FRM and coefficient decimation filters. The reconfigurable design outlined in \cite{haridas2017reconfigurable} is built on FRM and Farrow structure, in which the masking filters are incorporated as Farrow filters. In \cite{sudharman2019design}, two reconfigurable architectures based on non-maximally decimated filter bank (NMDFB) system and FRM are presented.
\begin{table}[]
\centering
\caption{Comparative analysis of the number of multipliers }
	\begin{tabular}{ll}
		\hline
		\textbf{Method} & \textbf{No. of Multipliers} \\ \hline
		Seperate FIR masking filter (9 standards)\cite{bindiya2012design} & 1249 \\
		VBW masking method  (9 standards)\cite{haridas2017reconfigurable} & 1168 \\
		Non uniform MDFT FB (9 standards)\cite{sakthivel2018low} & 918 \\
		Non maximally decimated FB method I (9 standards)\cite{sudharman2019design} & 614 \\
		Non maximally decimated FB method II (9 standards)\cite{sudharman2019design}& 586 \\
		Proposed method (9 standards) & 137 \\ \hline
	\end{tabular}{}
	\label{table:comparison}
\end{table}
\begin{figure}
	\centering
	\includegraphics[width=0.85\linewidth,height=4.25cm]{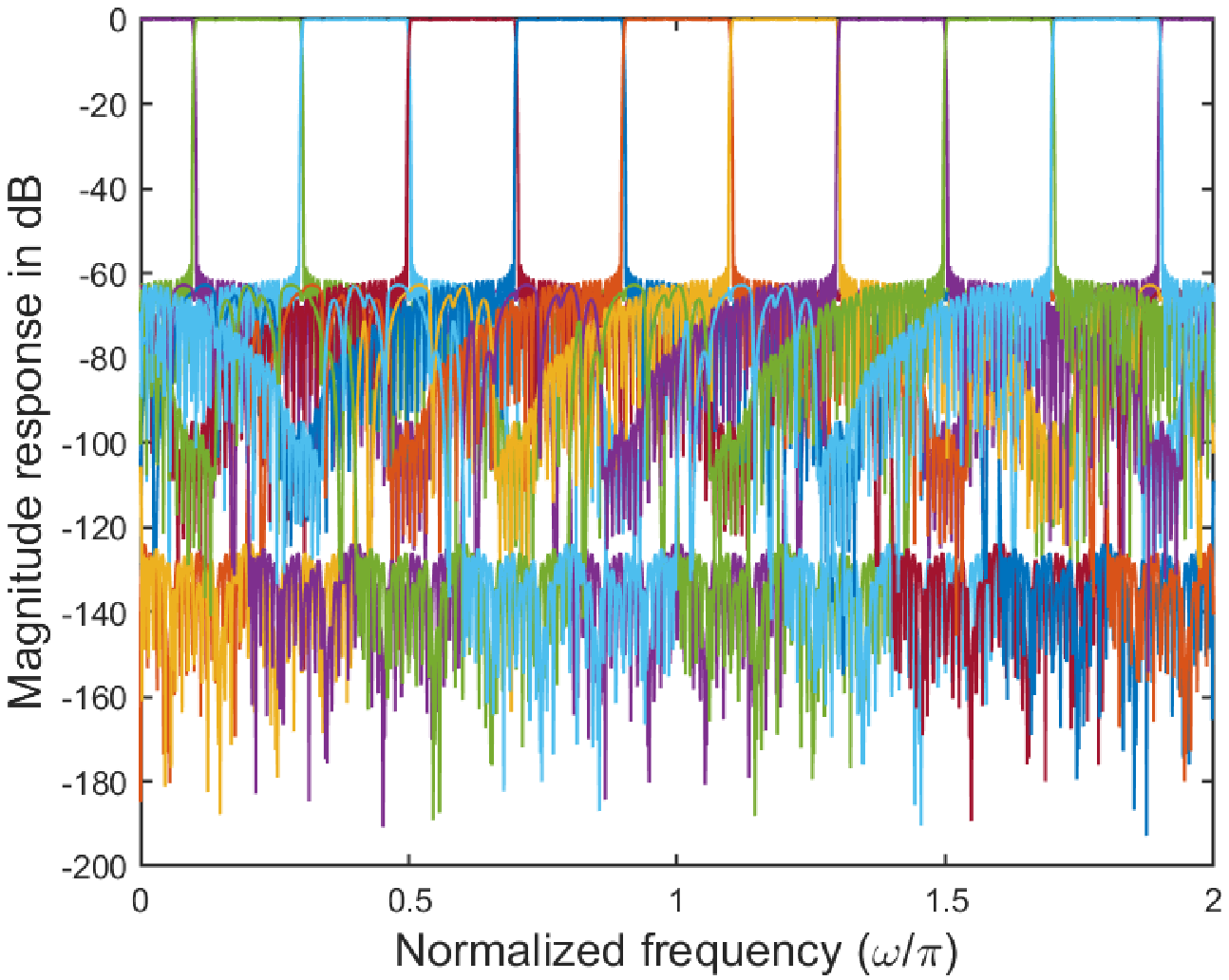}
	\caption{Uniform 10 channel ModFRM FB }
	\label{C10U}
\end{figure}
\begin{figure}
	\centering
	\includegraphics[width=0.85\linewidth,height=4.25cm]{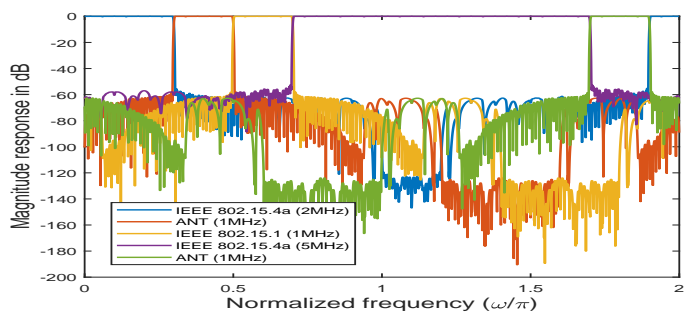}
	\caption{NU 5-channel ModFRM FB (for Bluetooth, ANT and Zigbee) derived from uniform 10-channel ModFRM FB.}
	\label{C10NUi}
\end{figure}
\par In this example, we will investigate the performance of reconfigurable
NU ModFRM FB as an
effective SDR channelizer. At first, uniform ModFRM FB of 8,10,12,16,32 channels are developed. The NU ModFRM FB are then generated by combining the specific nearby uniform ModFRM FB channels.
In this paper the same channels as described in Example III in \cite{haridas2017reconfigurable} are designed. This is to associate the complexity of the method proposed with the available schemes described in \cite{haridas2017reconfigurable} and \cite{sakthivel2018low}. The modal filter parameters and \textit{L} are chosen  according to the total number of channels to be achieved. Design specifications of the modal filter and the masking filters for the 8,12,16 and 32 channels are taken same as that in previous example. The modal filter specifications can be modified to satisfy the bandwidth requirements for the 10 channel as:\\
\textit{Pass band edge frequency :} $0.4\pi$\\
\textit{Stop band edge frequency :} $0.6\pi$\\ 
\textit{Maximum pass-band ripple :} 0.0065dB\\
\textit{Minimum stop-band attenuation :} 60dB \\
Here m is chosen as 1 and \textit{L} as 25 to satisfy (\ref{2pi}), (\ref{max_channels}) and (\ref{casei_ii}). Since the modal filter is a lower order filter, it is simple to reconfigure the modal filter.
\begin{figure}
	\centering
	\includegraphics[width=0.85\linewidth,height=4.25cm]{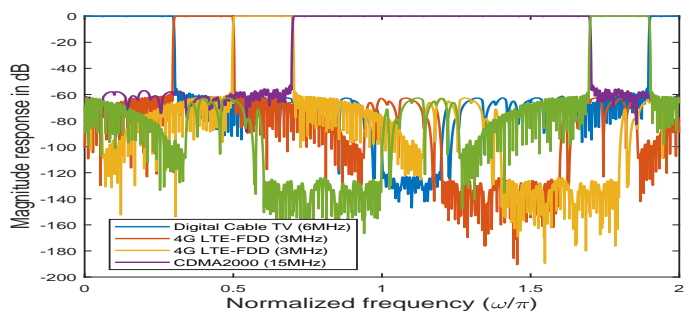}
	\caption{NU 5-channel ModFRM FB (Digital Cable TV, LTE and CDMA2000)derived from the same uniform10-channel ModFRM FB.}
		\label{C10NUii}
\end{figure}

\begin{figure}
	\centering
	\includegraphics[width=0.85\linewidth,height=4.25cm]{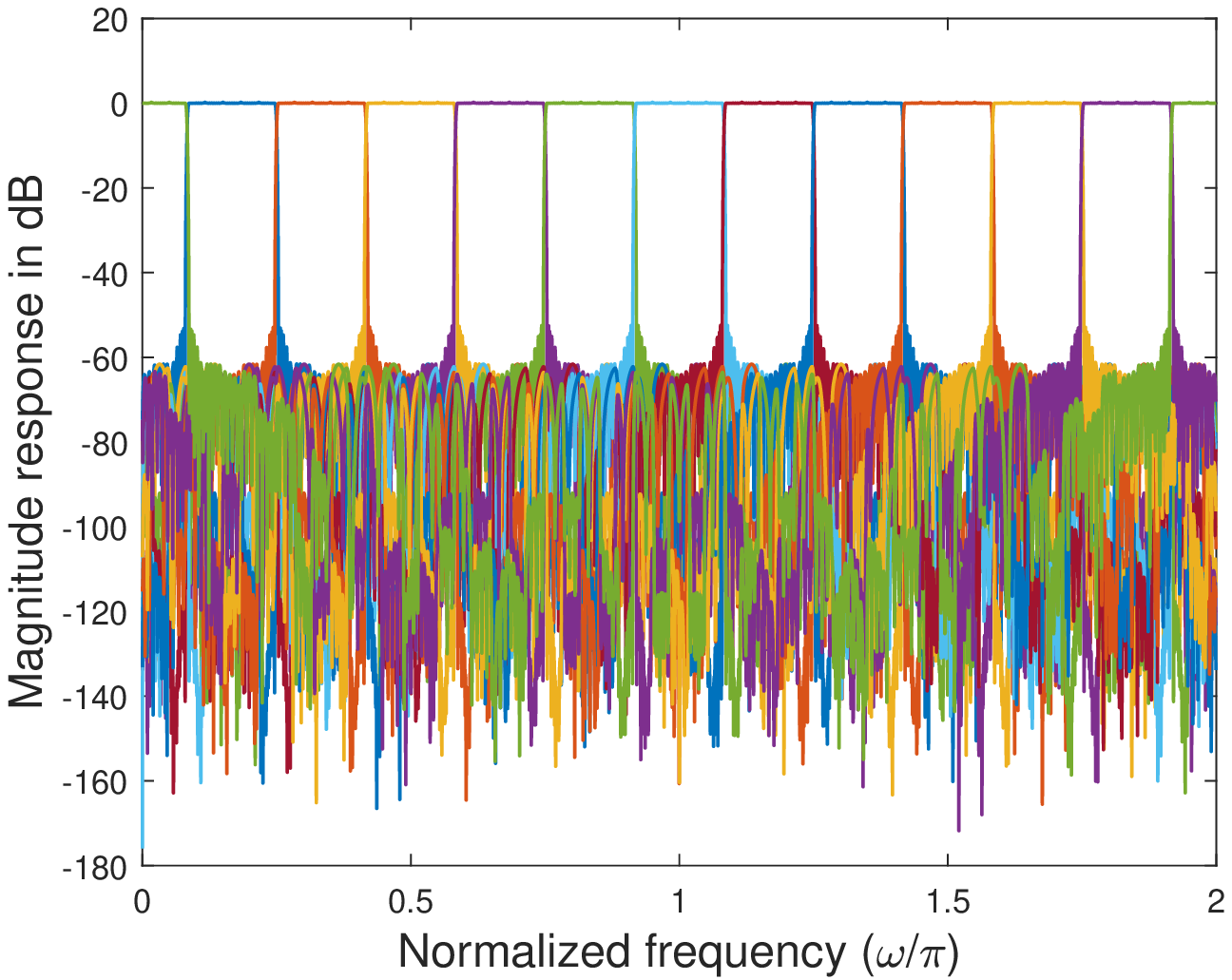}
	\caption{ Uniform 12-channel ModFRM FB.}
		\label{C12U}
\end{figure}

\begin{figure}
	\centering
	\includegraphics[width=0.85\linewidth,height=4.25cm]{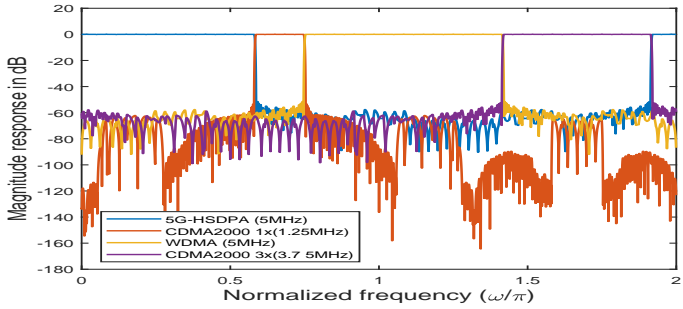}
	\caption{NU 4-channel ModFRM FB (for HSDPA, CDMA2000 and WCDMA) derived from uniform 12-channel ModFRM FB.}
		\label{C12NU}
\end{figure}
\par The magnitude response of the uniform 8-channel ModFRM FB is given in Fig. \ref{C8U}.
By combining the nearby channels with the channel allocation ($c_{0}$,$c_{1}$,$c_{2}$,$c_{3}$)=(2,1,3,2), 4 wireless CDMA2000 standards, as seen in Fig. \ref{C8NU4} is achieved. Uniform ModFRM FB are also configured for channels 10, 12, 16, and 32. The magnitude response of the 10-channel uniform ModFRM FB is given in Fig. \ref{C10U}. All filters are designed with the passband ripple and stopband attenuation as 0.0065dB and 60dB respectively. By combining the nearby channels with the channel allocation ($c_{0}$,$c_{1}$,$c_{2}$,$c_{3}$,$c_{4}$,)=(2,1,1,5,1), wireless standards like Bluetooth, ANT and Zigbee can be obtained, as in Fig. \ref{C10NUi}. The same 10-channel ModFRM FB can now be used to achieve varioous other wireless standards as shown in Fig. \ref{C10NUii}. This illustrates the possibility of reconfiguring the FB according to the requirements.

\begin{figure}
	\centering
	\includegraphics[width=0.85\linewidth,height=4.25cm]{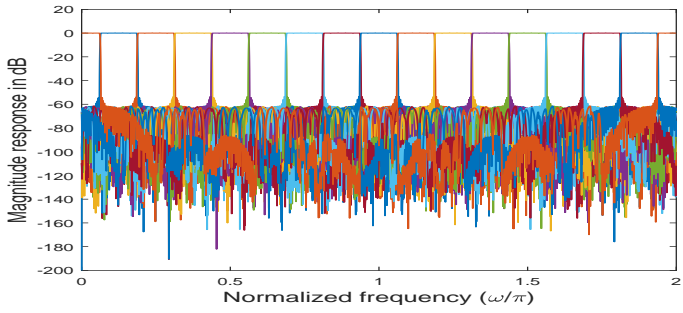}
	\caption{Uniform 16 channel ModFRM FB.}
		\label{C16u}
\end{figure}

\begin{figure}
	\centering
	\includegraphics[width=0.85\linewidth,height=4.25cm]{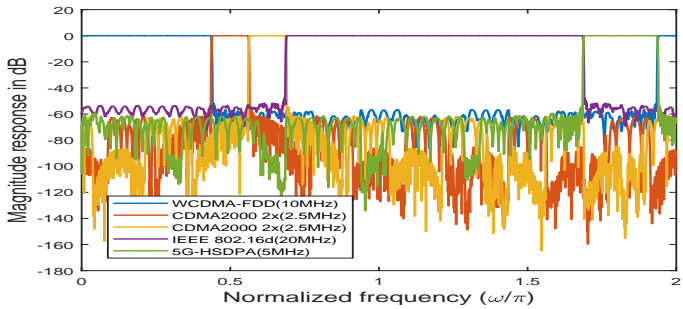}
	\caption{NU 5-channel ModFRM FB (for WCDMA, CDMA2000, FixedWiMAX and HSDPA) derived from uniform 16-channel ModFRM FB.}
		\label{C16NU}
\end{figure}

\begin{figure}
	\centering
	\includegraphics[width=0.85\linewidth,height=4.25cm]{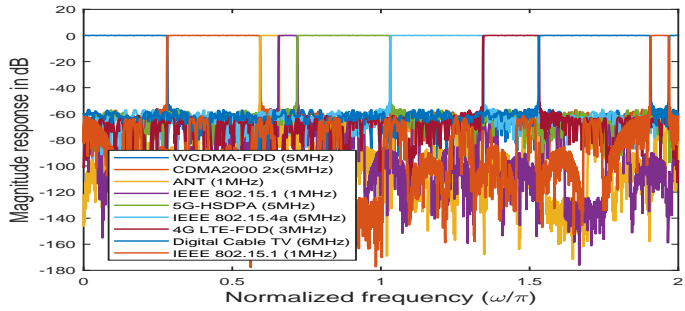}
	\caption{NU 9-channel ModFRM FB (for WCDMA, CDMA2000, Zigbee, ANT, Bluetooth, Digital Cable TV, LTE and HSDPA) derived from uniform 32-channel ModFRM FB.}
		\label{C32NU}
\end{figure}

The uniform 12-channel ModFRM FB is used to achieve a NU 4-channel ModFRM FB by merging the channels as ($c_{0}$,$c_{1}$,$c_{2}$,$c_{3}$)=(4,1,4,3) to generate multiple other wireless standards. The magnitude response of the uniform 12-channel ModFRM FB and the NU 4-channel ModFRM FB derived are given in Fig. \ref{C12U} and Fig. \ref{C12NU} respectively. The magnitude response of the uniform 16-channel ModFRM FB and the NU 5-channel ModFRM FB are given in Fig. \ref{C16u} and Fig. \ref{C16NU}, respectively. In this case, NU 5-channel ModFRM FB is derived from uniform 16-channel ModFRM FB to attain wireless standards such as CDMA2000, Fixed WiMAX, HSDPA and WCDMA. The adjacent channels are merged as ($c_{0}$,$c_{1}$,$c_{2}$,$c_{3}$,$c_{4}$)=(4,1,1,8,2). By merging the adjacent channels of the uniform 32-channel ModFRM FB in Fig. \ref{fig:M4L40C32} with the channel allocation ($c_{0}$,$c_{1}$,$c_{2}$,$c_{3}$,$c_{4}$,,$c_{5}$,$c_{6}$,$c_{7}$,$c_{8}$)=(5,5,1,1,5,5,3,6,1), NU 5-channel ModFRM FB can be derived as shown in Fig. \ref{C32NU}.

In Table \ref{table:comparison}, the complexity of the proposed SDR channelizer is contrasted with methods in \cite{haridas2017reconfigurable,sakthivel2018low,sudharman2019design} with respect to the number of multipliers. It is very clear from the table that hardware complexity of the proposed SDR channelizer is drastically less than the results in \cite{haridas2017reconfigurable,sakthivel2018low,sudharman2019design}. A 76.62\% reduction is observed in the number of multipliers when contrasted with  method-II in \cite{sudharman2019design}. Consequently, our design requires much less area, as the area consumed by the FB is directly proportionate to number of multipliers.

\section{Conclusion}
\par In this paper, a low complexity reconfigurable FB with narrow transition band is proposed. The modal filter and complementay filter in conventional FRM approach are replaced by a power complementary and linear phase FB in our proposal. With an alternate masking scheme, the multiple spectra replicas appearing in [0, $2\pi$] is fully utilized to generate more number of channels. Reconfigurability of filter bank is achieved by adjusting the interpolation values appropriately, to obtain more masking responses. Various ModFRM FB derived from the same modal filter are utilised for designing SDR channalizers that can cater to a broad variety of communication norms. The SDR channelizer designed using our proposed scheme had a lower implementation complexity in comparison to the existing methods in literature, for the same set of standards.

\bibliographystyle{IEEEtran}
\bibliography{ms}

\end{document}